\documentclass[a4paper,10pt]{aa}
\usepackage{epsfig}
\usepackage{fancyheadings}
\usepackage{amssymb}
\newcommand{\halfskip}{\vskip 0.5\baselineskip \noindent}
\newcommand{\be}{\halfskip \begin{equation}}
\newcommand{\ee}{\end{equation} \halfskip}
\newcommand{\nskip}{\vskip \baselineskip \noindent}
\begin{document}
\thesaurus{}
\authorrunning{E. van der Swaluw et al.}
\title{Pulsar wind nebulae in supernova remnants}
\subtitle{Spherically Symmetric Hydrodynamical Simulations}
\author{E. van der Swaluw \inst{1}, A. Achterberg \inst{1}, Y.A. Gallant
\inst{1,2} and G. T\'oth \inst{3}}
\institute{Astronomical Institute, Utrecht University, P.O.Box 80000, 3508 TA 
Utrecht, The Netherlands
\and
Osservatorio Astrofisico di Arcetri, Largo E. Fermi 5, 50125 Firenze, Italy
\and
Department of Atomic Physics, P\'azm\'any P\'eter s\'et\'any 1, 1117 Budapest, 
Hungary
}
\date{}
\maketitle
\begin{abstract}
A spherically symmetric model is presented for the interaction of a pulsar wind 
with the associated supernova remnant. This results in a pulsar wind nebula 
whose evolution is coupled to the evolution of the surrounding supernova remnant. 
This evolution can be divided in three stages.
The first stage is characterised by a supersonic expansion of the pulsar wind nebula 
into the freely expanding ejecta of the progenitor star. In the next stage the 
pulsar wind nebula is not steady; the pulsar wind nebula oscillates between
contraction and expansion due to interaction with the reverse shock of the supernova remnant: 
reverberations which propagate forward and backward in the remnant. 
After the reverberations of the reverse shock have almost completely vanished 
and the supernova remnant has relaxed to a Sedov solution, the 
expansion of the pulsar wind nebula proceeds subsonically. In this paper we 
present results from hydrodynamical simulations of a pulsar wind nebula 
through all these stages in its evolution. The simulations were carried out 
with the Versatile Advection Code.
\end{abstract}

\section{Introduction}
 
The explosion of a massive star at the end of its life as a supernova 
releases an amount of energy roughly equal to $10^{53}$ 
erg. Roughly 99\% of this energy is radiated away in the form of 
neutrinos as a result of the deleptonization of the $\sim 1 M_{\odot}$ stellar 
core as it collapses into a neutron star. The remaining (mechanical) energy 
of about $10^{51}$ erg is contained in the strong shock propagating through 
the stellar mantle, and ultimately 
drives the expansion of the supernova remnant (SNR).
 
In those cases where a rapidly rotating neutron star (pulsar) remains as a
`fossil' of the exploded star,
a pulsar wind, driven by the spindown luminosity of the 
pulsar, can be formed. 
The precise magnetospheric physics leading to such a pulsar
wind is not fully understood, but it is believed that a major fraction of 
the spin-down luminosity of the pulsar is converted into the mechanical
luminosity of such a wind.

The total rotational energy released by a Crab-like pulsar over its lifetime  
is of order $10^{49}-10^{50}$ erg. This is much less than the mechanical 
energy of $\sim 10^{51}$ erg driving the expansion of the SNR. 
Therefore, the {\em dynamical} influence of the pulsar wind on the 
global evolution of the supernova remnant itself will be small.
 From an observational point of view, however, the presence of a pulsar
wind can lead to a plerionic supernova remnant, where the emission at
radio wavelengths shows an extended, flat-spectrum central source associated 
with a Pulsar Wind Nebula (PWN). The best-known example of such a system
is the Crab Nebula, and about a half-dozen other PWNs are known unambiguously
around pulsars from radio surveys (e.g.\ Frail \& Scharringhausen, 1997). 
These surveys suggest that only young pulsars with a high spindown luminosity
produce observable PWNs at radio wavelengths.
At other than radio wavelengths, in particular X-rays,
there are about ten detections of PWN around pulsars both in our own galaxy
and in the large Magellanic cloud (LMC) (e.g.\ Helfand, 1998;
Table 1 of Chevalier, 2000).

The expansion of an isolated SNR into the general interstellar medium (ISM)
can be divided in four different stages (Woltjer, 1972; see also
Cioffi, 1990 for a review): the free 
expansion stage, the Sedov-Taylor stage, 
the pressure-driven snowplow stage and the momentum-conserving stage. 
In the models presented here we will only focus on the 
evolution of a pulsar wind nebula (PWN) during the first two stages of the 
SNR: the free expansion stage and the Sedov-Taylor stage. We will assume that
the pulsar is stationary at the center of the remnant, excluding such
cases as CTB80 (e.g.\ Strom, 1987; Hester \& Kulkarni, 1988), PSR1643-43 and 
PSR1706-44 (Frail et al., 1994), where the 
pulsar position is significantly excentric with respect to the SNR, 
presumably due to a large kick velocity of the
pulsar incurred at its birth, assuming of course
that SNR and pulsar are associated and we are not dealing with
a chance alignment of unrelated sources.
The case of a pulsar moving through the remnant with a significant
velocity will be treated in a later paper.

In this paper we compare (approximate) analytical expressions for the
expansion of a PWN in a supernova remnant with hydrodynamical simulations
carried out with the Versatile Advection  Code 
\footnote{See http://www.phys.uu.nl/\~{}toth/} (VAC). We confirm earlier 
analytical results 
(Reynolds \& Chevalier, 1984; Chevalier \& Fransson, 1992) which state that
the PWN is expanding supersonically when it is moving through the freely 
expanding ejecta of the SNR. Due to deceleration of the 
expanding SNR ejecta by the interstellar medium (ISM), a reverse shock 
propagates back to the center of the SNR (e.g. McKee, 1974 Cioffi et al. 1988)).  Due to the
presence of reverberations of the reverse shock in the SNR , the expansion of
the PWN goes through an unsteady phase when this
reverse shock hits the edge of the PNW.  After these reverberations  
have decayed, the expansion of the PWN through the ejecta of the SNR progenitor star 
continues subsonically with the PWN almost in pressure equilibrium with
the interior of the SNR.

This paper is organised as follows. In sections 2 and 3 we discuss 
the aforementioned two stages of the PWN/SNR system. In section 4 the
hydrodynamical simulations will be presented and compared with the analytical
expressions from section 2 and 3.

\section{Pulsar Wind Nebula in a freely expanding Supernova Remnant}

In the early stage of the evolution of a PWN, the SNR consists mostly of the
stellar ejecta expanding freely into the interstellar medium.
The PWN expands into these ejecta. The sound velocity in the interior of 
the SNR is much smaller than the expansion velocity of the PWN. 
The supersonic expansion of the PWN results in a 
shock propagating into the ejecta (see figure 1). 

An analytical equation for 
the radius of this shock can be derived for a constant spindown luminosity. 
Using this solution, the assumption of supersonic expansion will be checked 
{\em a posteriori}.
For simplicity we assume that the ejecta have a uniform density, 
\be
\label{Density}
	\rho_{\rm ej}(t) = \frac{3M_{\rm ej}}{4 \pi R_{\rm ej}^{3}} \; , 
\ee
and a linear velocity profile as a function of radius,
\be
	V_{\rm ej}(r) = \frac{r}{t} = 
	V_{0} \: \left( \frac{r}{R_{\rm ej}} \right)
	\; ,
\ee
with $R_{\rm ej}=V_0t$ the radius of the front of the ejecta. The
value of $V_0$ is determined by the requirement that the kinetic energy
of the ejecta equal the total mechanical energy $E_{0}$ of the SNR:
\be
\label{MechSNR}
	E_0 = \mbox{$\frac{1}{2}$} \: \rho_{\rm ej}(t) \: 
	\int_{0}^{V_{0}t}
	\: \left( \frac{r}{t} \right)^{2} \: 4\pi r^2 \: {\rm d}r =
	\mbox{$\frac{3}{10}$} \: M_{\rm ej} V_{0}^{2} \; .
\ee 
This yields:
\be
	V_0=\sqrt{\frac{10}{3} \: \frac{E_0}{M_{\rm ej}}}.
\ee
We assume that the stellar ejecta swept up by the strong 
shock which bounds the PWN collect in a thin shell, and that 
this material moves with the post-shock velocity. Neglecting the contribution 
of the thermal energy we can write the total (kinetic) energy of this shell, 
$E_{\rm shell}$, as:
\be
	E_{\rm shell}(t)=
	\mbox{$\frac{1}{2}$} \: M_{\rm sw}(t) \:
	\left(\mbox{$\frac{3}{4}$} \dot R_{\rm pwn}(t) + 
	\mbox{$\frac{1}{4}$} \frac{R_{\rm pwn}}{t} \right)^2 \; ,
\ee
where
\be
	M_{\rm sw}(t) \equiv M_{\rm ej} \: 
	\left({R_{\rm pwn}\over V_0 t}\right)^3
\ee
is the ejecta mass swept up by the pulsar wind nebula.
In deriving the post-shock velocity,
we assumed that the ejecta behave as an ideal non-relativistic gas with
adiabatic heat ratio $\gamma_{\rm ej} = 5/3$ and used the Rankine-Hugoniot jump
conditions for a strong shock.

The interior of the PWN is dominated by thermal
energy. The sound speed in a realistic PWN is close to the speed of light
$c$, while the expansion velocity is much less than $c$.
Perturbations in the pressure will be smoothed
out rapidly, on a sound crossing time $t_{\rm s} \sim R_{\rm pwn}/c$, much less than
the expansion time scale $t_{\rm exp} \sim R_{\rm pwn}/\dot{R}_{\rm pwn}$. 
Therefore, we can assume a nearly uniform pressure $P_{\rm pwn}$ in the PWN. 
The internal energy of the PWN then equals
\be
	E_{\rm pwn}={4\pi\over 3(\gamma_{\rm pwn} -1)} \:R_{\rm pwn}^3 
	\: P_{\rm pwn} \; .
\ee
Here we take $\gamma_{\rm pwn} =4/3$ because the pulsar wind nebula material is 
relativistically hot. The pressure of the interior of the PWN must roughly 
equal the pressure in the shocked ejecta
just downstream of the outer shock of the pulsar wind nebula 
at $R_{\rm pwn}$:
\be
	P_{\rm pwn}(t)={2\over{\gamma_{\rm ej} +1}} \: \rho_{\rm ej}(t) \: 
	\left(\dot R_{\rm pwn}(t) - \frac{R_{\rm pwn}}{t} \right)^2 \; .
\ee
Combining these relations yields:
\be
	E_{\rm pwn} = \frac{2M_{\rm sw}}{(\gamma_{\rm pwn}-1)(\gamma_{\rm ej} +1)} \:
	\:
	\left(\dot R_{\rm pwn}(t) - \frac{R_{\rm pwn}}{t} \right)^2 \; .
\ee
\nskip
Energy conservation for the PWN system reads:
\be
	E_{\rm shell}(t) + E_{\rm pwn}(t)= E_{\rm init}(t) + L_0 t \; .
\ee
Here $E_{\rm init}(t)$ is the kinetic energy which the swept-up ejecta would
have if they were freely expanding. This quantity can be obtained by integrating 
the kinetic energy density of ejecta in a sphere with radius $r < R_{\rm pwn}$ 
if there was no PWN. This yields:
\be
\label{Econs}
	E_{\rm init}(t)=E_0 \; \left(\frac{R_{\rm pwn}}{R_{\rm ej}} \right)^5.
\ee
After some algebra using the equations (\ref{Density})-(\ref{Econs}) one can 
obtain a power-law solution for the radius of the pulsar wind bubble:
\be
\label{Anpwn}
	R_{\rm pwn}(t)=C\left({L_0t\over E_0}\right)^{1/5}V_0t\propto t^{6/5},
\ee
where $C$ is a numerical constant of order unity:
\halfskip
\begin{eqnarray}
	&& C =\left({4\over{\displaystyle 15(\gamma_{\rm ej}+1)
	(\gamma_{\rm pwn}-1)}}+\frac{289}{240} \right)^{-1/5}
	\simeq 0.922 \\
	&& {\rm with}\quad \gamma_{\rm ej}
	= \frac{5}{3}, \quad \gamma_{\rm pwn} = \frac{4}{3}\nonumber
\end{eqnarray}
\halfskip

\begin{figure}
\begin{center}
\includegraphics[scale=0.5,angle=-90]{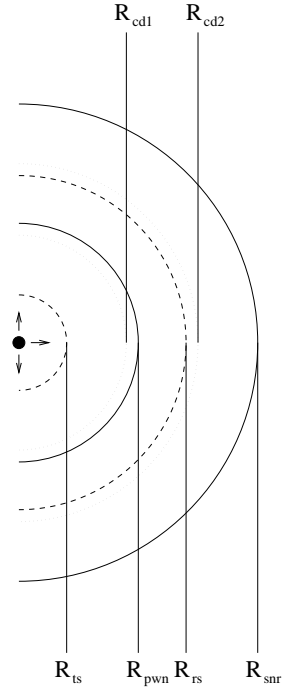}
\caption{Schematic representation of PWN in a freely expanding SNR. There are a 
total of four shocks and two contact discontinuities. From left to right one can 
see: the pulsar wind termination 
shock $R_{\rm ts}$ (dashed line), the first contact discontinuity $R_{\rm cd1}$ 
(dotted line) separating shocked pulsar wind material from shocked ejecta, 
the PWN shock $R_{\rm pwn}$ (solid line) bounding the PWN. For the SNR we 
have a reverse shock $R_{\rm rs}$ (dashed line), the second contact discontinuity 
$R_{\rm cd2}$ (dotted line) separating shocked ejecta from shocked ISM, and the 
SNR shock $R_{\rm snr}$ (solid line) which is the outer boundary of 
the PWN/SNR system.}
\end{center}
\end{figure}

Reynolds and Chevalier(1984) already obtained this 
$R(t)\propto t^{6/5}$ expansion law. It can easily be checked that the expansion
velocity obtained in this manner is indeed much larger than the sound velocity 
in the freely expanding supernova remnant.

\section{Pulsar Wind Nebula in a Sedov-Taylor remnant}
 
\subsection{Pulsar Wind Nebula expansion for a constant wind luminosity}

Towards the end of the free expansion stage a reverse shock is driven deep 
into the interior of the SNR. This reverse shock reheats the stellar ejecta, 
and as a result the sound velocity increases by a large factor. 
When the reverberations due to reflections of the reverse shock 
have almost completely dissipated, one can
approximate the interior of the SNR by using the analytical Sedov solution 
(\cite{Sedov}).

The interaction with the reverse shock influences the evolution of the pulsar 
wind nebula quite dramatically. Cioffi et al.\ (1988) have already shown in their 
1D simulation of a pure shell SNR that the reverse shock gives rise to all kinds
of sound waves and weak shocks traveling back and forth through the ejecta before the 
interior relaxes towards a Sedov solution. 
We will show that during the process of relaxation  
the radius of the pulsar wind nebula contracts and
expands due to reverberations of the reverse shock. Compression waves
are partly reflected and partly transmitted at the edge of the PWN. 
We will come back to this point when we discuss results from hydrodynamics simulations in 
section 4, which allow a more detailed picture of this process. 
 
In this Section we consider a fully relaxed Sedov SNR. 
The PWN expands subsonically into the remnant because the interior of the SNR has 
been re-heated by the reverse shock. 
For the case of a constant (mechanical) luminosity driving the pulsar wind 
an  analytical expression for the radius of the PWN can be easily obtained.
In this stage of the PWN evolution, we associate
its radius $R_{\rm pwn}$ with the contact discontinuity 
separating pulsar wind material from the ejecta of the progenitor star (see 
figure 2).

\begin{figure}
\begin{center}
\includegraphics[scale=0.5,angle=-90]{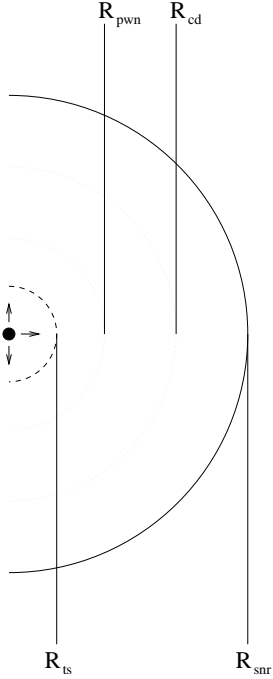}
\caption{Schematic representation of a PWN in a Sedov SNR. There are a total of 2 
shocks and 2 contact discontinuities. From left to right one can see: the
pulsar wind termination shock $R_{\rm ts}$ (dashed line), the first contact 
discontinuity $R_{\rm pwn}$ (dotted line) separating shocked pulsar wind 
material from shocked ejecta, bounding the PWN. Furthermore there is another 
contact discontinuity $R_{\rm cd}$ (dotted line) separating shocked ejecta 
from shocked ISM, and the SNR shock $R_{\rm snr}$ (solid line) which bounds the 
PWN/SNR system.}
\end{center}
\end{figure}

\ We first present an order-of-magnitude calculation which
leads to the correct power-law solution for the radius of the PWN. 
The assumption of  subsonic expansion implies approximate pressure 
equilibrium between the wind material and the stellar ejecta
at the edge the PWN. In the interior of the SNR the pressure scales as 
\be
	P_{\rm snr}\propto E_0/R_{\rm snr}^3 \; . 
\ee
On the other hand, the pressure in 
the interior of the PWN scales as 
\be
	P_{\rm pwn}\propto L_0t/R_{\rm pwn}^3 \; ,
\ee
with $L_{0}$ the mechanical luminosity driving the wind.	 
Pressure equilibrium at the contact discontinuity at 
$R_{\rm pwn}$ implies the following relation for the radius of the 
PWN as a function of time:
\be
\label{pwnexp}
	R_{\rm pwn}(t) = {\bar C}
	\left({L_0t\over E_0}\right)^{1/3}R_{\rm snr}(t)\propto t^{11/15},
\ee
with the constant of proportionality $\bar{C}$ to be determined below.

A more detailed derivation uses the first law of 
thermodynamics, assuming once again a constant energy 
input $L_{0}$ into the PWN by the pulsar-driven wind: 
\be
	{\rm d}E_{\rm th}=L_{0} \: {\rm d}t - P_{\rm i} \: {\rm d}{\cal V}_{\rm pwn}.
\ee
\noindent
Here $E_{\rm th}$ is the thermal energy of the PWN, 
$P_{\rm i}$ its internal pressure, and ${\cal V}_{\rm pwn}$ its volume. 
This yields the following equation describing the energy balance of a slowly
expanding PWN:
\be
\label{ThermD}
	{{\rm d}\over {\rm dt}}\left({4\pi\over 3}{P_{\rm i}R_{\rm pwn}^3\over 
	(\gamma_{\rm pwn} -1)}\right)
	=  L_0- 4\pi R_{\rm pwn}^2 \: P_{\rm i}\: 
	\left( \frac {{\rm d} {R}_{\rm pwn}}{{\rm d} t} \right) \; ,
\ee
\noindent 
or equivalently
\be
	{{\rm d}\over {\rm dt}}\left({4\pi\over 3}
	{\gamma_{\rm pwn} P_{\rm i}R_{\rm pwn}^3\over(\gamma_{\rm pwn} -1)}\right)
	= L_0 + {4\pi\over 3} \: R_{\rm pwn}^3 \: \left(
	\frac{{\rm d} {P}_{\rm i}}{{\rm d} t} \right) \; .
\ee  
This equation has a power-law solution for $R_{\rm pwn}(t)$
provided the internal pressure $P_{\rm i}(t)$ in the SNR behaves as 
a power-law in time so that the relation
\halfskip 
\be
	R_{\rm pwn}^3 \: \left( \frac{{\rm d} P_{\rm i}}{{\rm d} t} \right)
	 \; = \; \mbox{constant}
\ee
\halfskip
can be satisfied. For a Sedov SNR expanding into a uniform ISM one has
$P_{\rm i} \propto t^{-6/5}$ and one finds:
\be
	R_{\rm pwn}(t)= D \left({L_{0} t\over P_{\rm i}(t)}\right)^{1/3}
	\propto t^{11/15} \; , 
\ee
where
\halfskip
\be
	D = \left[ \frac{4\pi}{3} \:  
	\left({\gamma_{\rm pwn}\over\gamma_{\rm pwn} -1}+\frac{6}{5} \:
	\right) \right]^{- 1/3}
	\; .
\ee
\halfskip	 
If  $R_{\rm pwn} \ll R_{\rm snr}$ we can use the central pressure from the 
Sedov solution with $\gamma_{\rm ism} =5/3$ for the interior pressure in the SNR
which confines the PWN (e.g. Shu, 1992):
\be
\label{Shurel}
	P_{\rm i}(t)=P_{\rm snr}(t) \; \simeq \;  0.074\:
	\left( {E_0\over R_{\rm snr}^3} \right)
	\propto t^{-6/5} \; .
\ee
We find the same result for $R_{\rm pwn}(t)$ as in the order-of-magnitude 
calculation (Eqn. \ref{pwnexp}), determining the constant in that expression
as $\bar C\simeq 0.954$  for a non-relativistic fluid ($\gamma_{\rm pwn} = 5/3$) 
and $\bar C\simeq 0.851$ for a relativistic fluid ($\gamma_{\rm pwn} = 4/3$). 
By comparing the sound speed with the expansion velocity at the edge of the 
PWN, we confirm that the expansion remains subsonic.

An alternative derivation of the PWN expansion law uses the 
Kennel-Coroniti model for a highly relativistic pulsar-driven pair wind. 
This wind is terminated by a strong MHD shock which decelerates the fluid 
to a nonrelativistic expansion speed (Rees \& Gunn, 1974).
Kennel \& Coroniti (1984, hereafter K\&C) constructed a steady, 
spherically symmetric MHD model for the Crab nebula which includes these 
characteristics. We use their model in the hydrodynamical limit by 
considering the case 
\be
	\sigma \equiv \frac{\mbox{Poynting flux}}{\mbox{particle energy flux}}
	\rightarrow 0 \; . 
\ee
K\&C assume a constant wind luminosity,
\be
\label{Lzero}
	L_0 = 4\pi n_1 \: \Gamma_1 u_1 R^2_{\rm ts} \:  mc^2 \approx
	4 \pi n_{1} \: \Gamma_{1}^{2} R^{2}_{\rm ts} \: mc^{3} \; ,
\ee
\noindent 
where $n_{1}$ is the proper density just in front of the termination shock, 
$u = \Gamma_{1} \: v \approx \Gamma_{1} c$ 
is the radial four-speed of the wind and $R_{\rm ts}$ is the distance from the pulsar
to the termination shock. Because the wind is assumed to consist solely of a 
positronic plasma, $m$ is the electron mass. The pulsar wind is highly 
relativistic ($\Gamma_{1} \gg 1$) and the thermal and rest energy of the particles 
can be neglected compared with the bulk kinetic energy. The total number of particles 
emitted into the PWN then equals:
\be
\label{Ntotal}
	N(t) = {L_0 t\over\Gamma_1 \: mc^2} \; .
\ee
It is believed that the bulk Lorentz factor $\Gamma_1 \approx 10^6$, 
but we will see that for purposes of the PWN evolution
its precise value is not important, because it cancels in the final result.

\begin{figure}
\begin{center}
\includegraphics[scale=0.3,angle=-90]{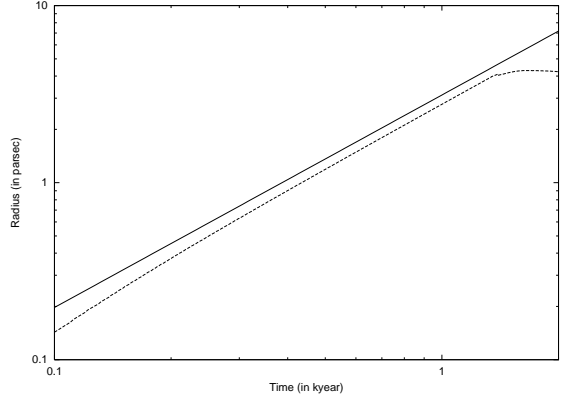}
\caption{Comparison between results from numerical simulations and
analytical result for the radius of the PWN, i.e. equation (\ref{Anpwn}). The dashed
line indicates the radius for the PWN obtained from numerical simulations. 
The solid line corresponds to equation (\ref{Anpwn}) with $C \simeq 0.941$, as appropriate for
$\gamma_{\rm pwn} =5/3$.  The different  physical parameters are as indicated
in Table 1 (Simulation 1). 
The injected mass of the 
pulsar wind has been chosen in such a way that the termination velocity of the 
pulsar wind equals the speed of light. One can see that in the simulation 
the radius $R_{\rm pwn}$ is about 10 \% smaller than predicted by the analytical result,
but the power-law behaviour $R_{\rm pwn} \propto t^{6/5}$ is correctly reproduced.} 
\end{center}
\end{figure}

\noindent 
After the termination of the cold wind by a strong standing shock at 
some radius $R_{\rm ts}$, the wind flow is subsonic, with a sound speed 
close to the speed of light. 
Assuming that the shock radius $R_{\rm ts}$ is much smaller than
the radius $R_{\rm pwn}$ of the PWN, and assuming a uniform density $n_{2}$
and uniform pressure $P_{2}$ inside the PWN, particle conservation implies
\be
\label{conservation}
	{4\pi\over 3} \: n_{2} \: R_{\rm pwn}^3(t)  \: = \: 
	{L_0 t\over\Gamma_1 \: mc^2} \; .
\ee
\noindent 
 From the K\&C model we take the following relationships, 
valid at the strong relativistic termination shock at the inner edge of
the PWN in the hydrodynamical limit:
\be
\label{pressurePWN}
	n_2  = \sqrt{8} n_1 \Gamma_1 \; \; , \; \; 
	P_2 = {2\over 3} n_1 \Gamma_1^2 \: mc^2
	\approx \frac{L_{0}}{6 \pi R_{\rm ts}^{2} \: c} \; .	
\ee
The subscripts 1 and 2 label upstream and downstream parameters on either
side of the termination shock. Using these jump conditions together with 
equations (\ref{Lzero}) and (\ref{Ntotal}) 
we can express $R_{\rm pwn}(t)$ as a function of $R_{\rm ts}(t)$:
\be
\label{radrelat}
	R_{\rm pwn}(t) = \left({3ct\over\sqrt{8}}\right)^{1/3}
	\: R_{\rm ts}^{2/3}(t).
\ee
The pressure inside the PWN is nearly uniform. At the termination shock 
(inner edge of the PWN) it must  equal the downstream pressure $P_{2}$. 
At the outer edge of the PWN this pressure must approximately
equal the pressure $P_{\rm snr}(t)$ at the center of the 
SNR as given by Eqn. (\ref{Shurel}).
Using (\ref{pressurePWN}) and (\ref{radrelat}) the inner and outer boundary
conditions imply the following relation for the termination shock radius:
\be
\label{termrad}
	R_{\rm ts}(t)\simeq 0.847 \: 
	\left({L_{0}\over E_0 \: c}\right)^{1/2}R_{\rm snr}^{3/2}(t) \; .
\ee   
\noindent 
It is now straightforward to obtain the radius of the PWN from (\ref{radrelat})
and (\ref{termrad}). The resulting expression for $R_{\rm pwn}$ satisfies 
equation (\ref{pwnexp}) 
with $\bar C\simeq 0.911$. This derivation based on (ram) pressure balance at the
inner and outer edges of the pulsar wind nebula confirms our earlier result obtained
from overall energy conservation. 

\subsection{Pulsar Wind Nebula expansion for varying wind luminosity}

The constant wind luminosity assumption is not very realistic by the time
the effects of the reverse shock and its associated reverberations have vanished. 
The spin-down luminosity of the pulsar is more realistically described by the 
luminosity evolution from a rotating magnetic dipole model:
\be
\label{pulsarlum}
	L(t)=\frac{L_0}{\displaystyle \left(1+{t\over\tau}\right)^2}.
\ee
Therefore we now consider the more realistic case of a time-dependent luminosity 
given by (\ref{pulsarlum}).
The energy balance equation for the PWN reads:
\halfskip
\begin{eqnarray}
\label{SemiPWN}
	&& {{\rm d}\over {\rm dt}}\left({4\pi\over 3}
	{P_{\rm i}R_{\rm pwn}^3\over (\gamma_{\rm pwn} -1)}\right) = 
	{L_0\over\left( 1+{t/\tau}\right)^2} - \nonumber \\ 
	&& 4\pi R_{\rm pwn}^2 \: P_{\rm i} \: 
	\left( \frac{{\rm d} {R}_{\rm pwn}}{{\rm d} t} \right)
	\; .
\end{eqnarray}
\halfskip
We solve this equation numerically using a fourth-order Runge-Kutta 
method (e.g. Press et al., 1992). As an initial condition we 
take the radius of the PWN equal to zero at the start of the evolution, 
neglecting the initial stage when the PWN is expanding supersonically. For the
pressure $P_{\rm i}$,  we use the pressure at the center of the Sedov SNR 
(\ref{Shurel}). We find 
that the solution for $R_{\rm pwn}$ converges to $R_{\rm pwn}\propto t^{0.3}$
on a time scale much larger than the typical time scale for the reverse
shock to hit the edge of the PWN. Figure 9 shows this semi-analytical result 
together with results from hydrodynamical simulations. For the semi-analytical 
equation we use $\gamma_{\rm pwn} =5/3$, because the hydrodynamics code also 
uses this value (see section 4.1 below).

\section{Numerical simulations}

\subsection{Method}

Our simulations were performed using the Versatile Advection Code 
(VAC, T\'oth 1996) which can integrate the equations of gas dynamics
in a conservative form in 1, 2 or 3 dimensions. We used the TVD-MUSCL
scheme with a Roe-type approximate Riemann solver
from the numerical algorithms available in VAC 
(T\'oth and Odstr\v cil, 1996); a discussion of this and other schemes 
for numerical hydrodynamics can be found in LeVeque (1998). In this paper 
our calculations are limited to spherically symmetric flows.

\par\noindent 
We use a uniform grid with a grid spacing chosen sufficiently fine 
to resolve both the shocks  
inside the PWN and the larger-scale shocks associated with the SNR. 
Table 1 gives the physical scale associated with the grid size 
for the simulations presented here. An expanding SNR is created by
impulsively releasing the mechanical energy of the SN explosion in the first 
few grid cells.  The thermal energy and mass deposited there lead to 
freely expanding ejecta with a nearly uniform density, 
and a linear velocity profile as a function of radius.  

\begin{figure}
\begin{center}
\includegraphics[scale=0.45,angle=0]{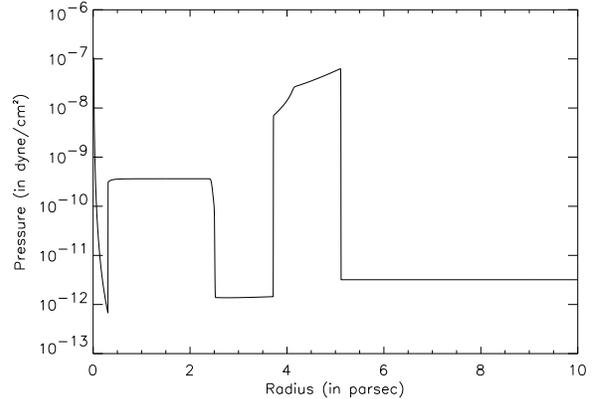}
\caption{Pressure profile of the PWN/SNR system as a function of radius,
at time $t=1000$ years after the SN explosion.  Physical parameters are
as indicated in Table 1 (Simulation 2).
Moving outwards in radius one can see the wind termination shock, 
the shock bounding the PWN, 
the reverse shock of the SNR and the shock bounding the SNR. 
The interior of the PWN is nearly isobaric. 
There is a sudden increase in pressure of the ejecta  behind the SNR reverse 
shock.}
\end{center}
\end{figure}

\begin{figure}
\begin{center}
\includegraphics[scale=0.45,angle=0]{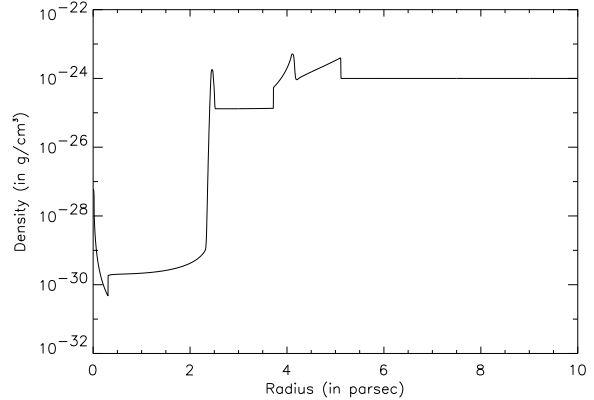}
\caption{Density profile for the same PWN/SNR system as in figure 4.}
\end{center}
\end{figure}

\noindent A realistic shocked pulsar wind is presumably highly relativistic 
with an adiabatic heat ratio $\gamma_{\rm pwn} = 4/3$. 
The (shocked) stellar ejecta 
on the other hand are non - relativistic with $\gamma_{\rm ej} = 5/3$. 

The VAC code does not currently include relativistic hydrodynamics. 
Therefore, the best approach available to us is to keep
$\gamma_{\rm pwn} = 5/3$, but to take a 
luminosity for the pulsar wind, $L(t)$, and an associated mass injection, 
$\dot M_{\rm pw}(t)$, such that the terminal velocity obtained from these 
two parameters,
\be
	v_\infty =\sqrt{2L(t)/\dot M_{\rm ej}(t)} \; , 
\ee
roughly equals the speed of light. 
Since the pulsar wind material downstream of the termination shock moves with
only a mildly relativistic bulk speed we expect our results to be qualitatively 
correct. Thermal energy and mass are deposited continuously in a small
volume as a source for the wind.
The hydrodynamics code then develops a steady 
wind reaching the terminal velocity $v_\infty$ well before the (cold) wind 
is terminated by the standing termination shock.   

We trace the total mass injected into the PWN by the pulsar wind in order 
to determine the radius of the contact discontinuity which separates the
pulsar wind material from 
the SN ejecta ($R_{\rm cd1}$ in figure 1 and $R_{\rm pwn}$ in figure 2). 
We also determine the position of the shock bounding the PWN during the stage
of supersonic expansion.
This enables us to compare the numerical results with the 
analytical expressions derived in sections 2 and 3 for the PWN radius.

\begin{figure*}
\begin{center}
\includegraphics[scale=0.7,angle=270]{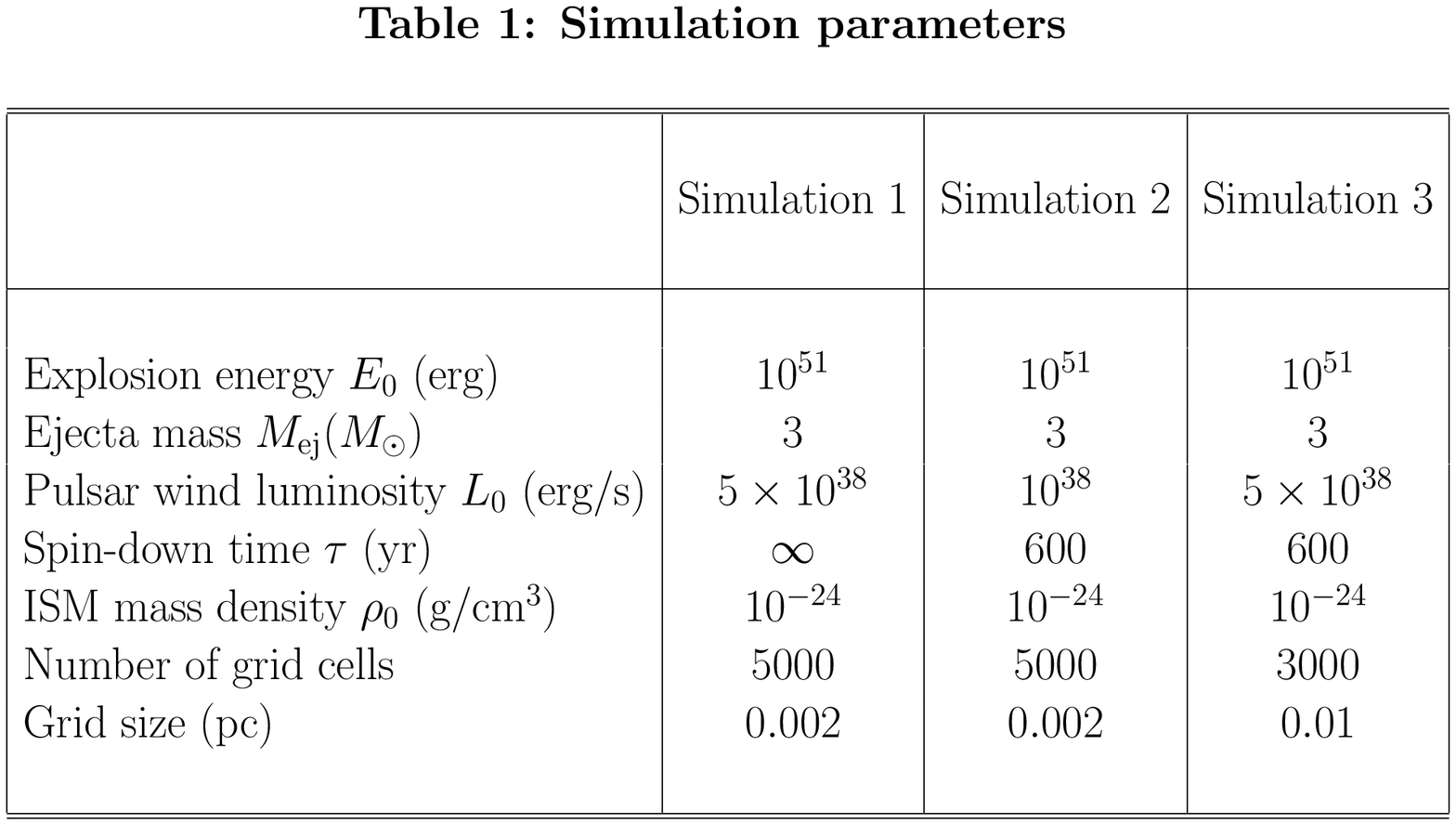}
\end{center}
\end{figure*}

\begin{figure}
\begin{center}
\includegraphics[scale=0.45,angle=0]{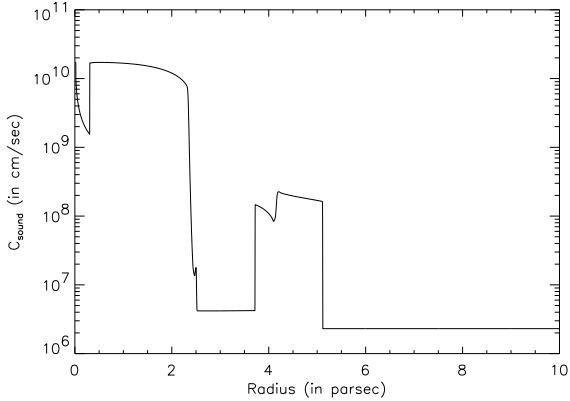}
\caption{Sound velocity profile as a function of radius, for the same case
as in figures 4 and 5.
Because of the Rankine-Hugoniot jump conditions at the wind termination shock,
the sound velocity of the shocked wind material
in the PWN bubble is close to the speed of light. Behind the contact discontinuity,
where the bubble consists of swept-up ejecta, the sound speed has a smaller value.}
\end{center}
\end{figure}

\begin{figure}
\begin{center}
\includegraphics[scale=0.45,angle=0]{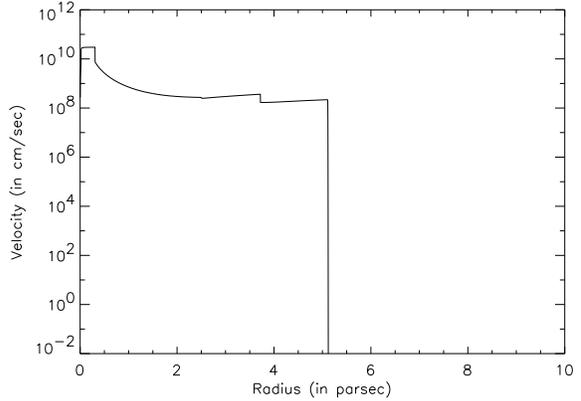}
\caption{Velocity profile for the PWN/SNR system with the same parameters. 
The terminal wind velocity, $v_\infty$, is close to the speed of light. 
The large jump in the velocity at a radius $\sim 4$ pc is the reverse shock
which is still propagating forwards in the laboratory frame. 
The velocity jump of the PWN shock at radius $\sim 2.5$ pc is much smaller.}
\end{center}
\end{figure}

As a test of the code  we have calculated a pulsar wind driven by a constant 
luminosity $L_{0}$ (Simulation 1 in Table 1).  
We let the PWN evolve until the reverse shock propagating in the SNR
hits its outer edge. Figure 3 shows the radius of the shock of the PWN 
together with the analytical 
equation (\ref{pwnexp}). We take $\gamma_{\rm pwn} =5/3$ in the analytical expressions 
for comparison with the numerical results. Although the analytical result of 
Eqn. (\ref{pwnexp}) 
is not reproduced exactly (the radius
is about 10\% smaller), the power-law expansion law 
$R_{\rm pwn} \propto t^{6/5}$ {\em is} reproduced. 
As we will show in section 4.2, the pressure inside the
bubble is larger than the one used to derive the equation, explaining the
difference between the analytical and numerical results.
 
\subsection{Evolution of the PWN-SNR system into the
Sedov phase}

Our simulations of the evolution of a pulsar wind nebula inside a supernova 
remnant employ the parameters listed in Table 1. 

In the early stage of its evolution the 
PWN is bounded by a strong shock propagating through the ejecta of the 
progenitor star. In figures 4--7 one can clearly identify the four 
shocks indicated schematically in figure 1. 
Moving outward in radius one first encounters the pulsar wind termination shock;
this termination shock is followed by the PWN shock. 
In the sound velocity profile of figure 6 one can see a large jump between 
these two shocks: 
the contact discontinuity separating shocked pulsar wind material 
from shocked ejecta. 
Further outward one encounters the SNR reverse shock, 
which at this stage of the SNR evolution is still moving outwards 
from the point of view of a stationary outside observer. 
The whole PWN-SNR system is bounded by the SNR blast wave.

Figure 9 shows the evolution of the contact discontinuity radius $R_{\rm cd}$,
which can be identified with the radius of the PWN in the subsonic 
expansion stage. One can clearly see the moment at $t\simeq 1.75$
kyr when the reverse shock hits the edge of the PWN: the expansion becomes 
unsteady with the PWN contracting and expanding due to the interaction with
the pressure pulses associated with the reverberations of the 
reverse shock. When these reverberations have almost 
dissipated the expansion of the PWN relaxes to a steady subsonic
expansion. In this stage, we can fit the radius of the PWN obtained 
from the simulations with the \mbox{(semi-)}analytical solution obtained 
from a numerical integration of equation (\ref{ThermD}), as shown in this 
figure. 

The interaction of the PWN with the reverse shock and the associated 
reverberations is quite complicated. We will therefore describe this process 
in more detail.

\begin{figure}
\begin{center}
\includegraphics[scale=0.3,angle=-90]{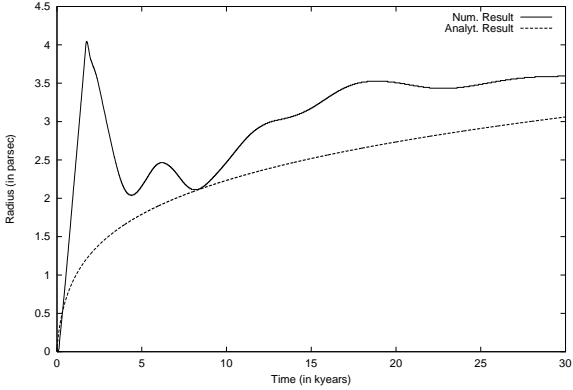}
\caption{The radius of the PWN contact discontinuity as a function of time
(solid line). We compare with the semi-analytical solution from equation
(\ref{SemiPWN}) (dashed line). Here one can see that the 
expansion of the PWN is unsteady, due to the reverberations of the reverse
shock. This simulation was done with the parameters listed in Table 1 
(Simulation 3).}
\end{center}
\end{figure}

\begin{figure}
\begin{center}
\includegraphics[scale=0.45,angle=-90]{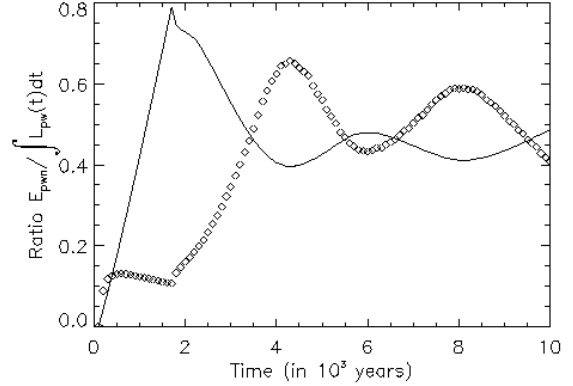}
\caption{The radius of the PWN as a function of time, together with the ratio of
the total energy in the PWN bubble with respect to the total energy input by
the pulsar wind. The solid line represents the radius of the contact discontinuity
of the PWN (in arbitrary units), the open squares represent the aforementioned ratio
of energy. This simulation was done with the parameters as listed in Table 1
(Simulation 3).}
\end{center}
\end{figure}

\subsection{The influence of reverse-shock reverberations}

The reverse shock initially encounters the PWN in its supersonic expansion stage.
After the collision between the reverse shock and the PWN shock a 
reflected shock propagates back towards the outer
(Sedov-Taylor) blast wave of the SNR. 
A transmitted shock propagates into the shocked ejecta  inside the PWN. 
When this shock hits the contact discontinuity bounding the pulsar wind material 
a similar reflection/transmission event occurs: a shock moves radially 
outwards, and a compression wave moves into the pulsar wind material. The 
latter wave is rapidly dissipated in the pulsar wind bubble because of the 
high sound speed in the shocked pulsar wind. 
After a few sound crossing times the pulsar wind bubble contracts adiabatically
in response to the pressure increase inside the SNR. After this contraction 
it regains pressure equilibrium with the surrounding SNR and the PWN expands
subsonically henceforth. This chain of events can be clearly seen in figure 8
where we plot the radius of the PWN. The whole process takes a 
time comparable with the duration of the initial
supersonic expansion stage.

\subsection{Subsonic expansion stage}

When the PWN has more or less relaxed 
to a steady subsonic expansion the PWN has gained energy 
as a result of the interaction with the reverse shock. Consequently, 
the radius of the PWN is roughly 20\% larger than the value
predicted by the semi-analytical solution obtained from Eqn. (\ref{ThermD}) in Section 3.2. 
In figure 9 we show the ratio between the (mostly thermal) energy of the pulsar
wind bubble, i.e.\ the part of the PWN that consists of shocked pulsar wind
material, and the total mechanical energy deposited by the pulsar. 
One can clearly see the increase in the energy content
of the pulsar wind bubble. A large fraction of the energy deposited by the pulsar
wind in the stage when the expansion is supersonic is contained in the kinetic 
energy of the shocked stellar ejecta in the PWN shell. When the reverse SNR 
shock is interacting with the PWN bubble, energy is apparently 
transferred from this thin shell to the interior of the bubble 
through the dissipation of the waves transmitted into the bubble.

\section{Conclusions and discussion}

We have considered a spherically symmetric PWN/SNR system in the 
early and middle stages of its evolution, 
well before cooling of the SNR shell becomes dynamically important and 
before a significant disruption of spherical symmetry due to a possible 
(large) kick velocity of the  pulsar can take place. The expansion of the 
PWN is coupled with the dynamics of the expanding SNR,
leading to two distinct evolutionary stages separated by an unsteady
transition phase:

\begin{itemize}

\item
When the PWN is surrounded by the freely expanding ejecta of 
the SNR, the expansion of the PWN is supersonic. In this stage 
the pressure in the interior of the PWN bubble is slightly larger than 
one would expect from ram pressure of the surrounding ejecta alone, using the
Rankine-Hugoniot relations at the PWN shock. This is 
due to the thin shell of shocked, swept-up ejecta which needs to be 
accelerated by the outward force due to the interior pressure of the PWN.

\item
This stage of supersonic expansion is ultimately followed by a 
subsonic expansion of the PWN. This happens after the reverse shock has
encountered the shock of the PWN.

\item
The transition between these two stages is unsteady due to the 
interaction of the PWN with the reverse shock and its associated 
reverberations. From the hydrodynamical simulations we see that the 
time scale for adjustment to the pressure of the surrounding SNR is determined 
by the sound speed of the ejecta shocked by the PWN in the first stage. 

\end{itemize}

Two of the prototypes of plerionic SNRs are the Crab Nebula and 3C58. In the
Crab, there is a decrease in the radio flux of 0.167 $\pm$ 0.015 
\% yr$^{-1}$ (Aller \& Reynolds 1985).
By contrast, 3C58 shows an increase in its flux density at radio frequencies
between 1967 and 1986 (Green 1987). This increase might be the result of the 
reverse shock which has encountered the PWN shock around 3C58; the PWN is 
being compressed and therefore the flux density is going up.

Our numerical simulations are different from the results presented by
Jun (1998). This author concentrates on the details of the PWN in the 
supersonic expansion stage, and in
particular on the formation of Rayleigh-Taylor fingers in his two-dimensional
simulations. Our simulations include the whole supernova remnant, but can not
address the development of Rayleigh-Taylor instabilities due to our assumption 
of spherical symmetry.

In future work we will discuss 
how these results change when the influence of a significant kick velocity 
of the pulsar is taken into account. If this is taken into account, the model
presented here will lose its validity at a certain time: one can calculate when
the motion of a pulsar will become supersonic in a Sedov stage. One can show
that this will happen when the pulsar is about halfway from the explosion
center to the edge of the SNR: a bow shock is
expected to result from this and clearly the model presented here will break
down. Observationally there is evidence that this is the case for the pulsar
associated with the SNR W44.

\begin{acknowledgements} 
\noindent The authors would like to thank L.Drury for discussions on the subject,
T.Downes for discussing several numerical techniques, D.Frail for clarifying the
nature of W44, and R.Keppens for his assistance with the code VAC. 
The Versatile Advection Code has been developed as part of the project on
``Parallel Computational Magneto-Fluid Dynamics'', funded by the Dutch
Science Foundation (NWO) from 1994 to 1997.  Y.A.G. acknowledges support from 
the Netherlands Foundation for Research
in Astronomy (ASTRON project 781--76--014) and the Italian Ministry of
Universities and Research (grant Cofin--99--02--02).
G.T. is currently supported by a postdoctoral 
fellowship (D~25519) from the Hungarian Science Foundation (OTKA).
\end{acknowledgements}

\end{document}